\documentclass[aps,pra,twocolumn,groupedaddress]{revtex4-2}
\usepackage{graphicx,color}
\usepackage{dcolumn}
\usepackage{amssymb,amsmath,amsbsy,bm,amscd}
\usepackage[colorlinks=true,citecolor=blue]{hyperref}
\usepackage[table]{xcolor}
\usepackage{physics}

\begin{document}

\title{Race for Quantum Advantage using Random Circuit Sampling}
\author{Sangchul Oh}
\affiliation{Department of Chemistry, Department of Physics and Astronomy, and Purdue Quantum Science and
    Engineering Institute, Purdue University, West Lafayette, IN, USA}
\author{Sabre Kais}
\email{kais@purdue.edu}
\affiliation{Department of Chemistry, Department of Physics and Astronomy, and Purdue Quantum Science and
    Engineering Institute, Purdue University, West Lafayette, IN, USA}
\date{\today}

\date{\today}

\begin{abstract}
Random circuit sampling, the task to sample bit strings from a random unitary operator, has been performed to 
demonstrate quantum advantage on the Sycamore quantum processor with 53 qubits and on the Zuchongzhi quantum 
processor with 56 and 61 qubits. Recently, it has been claimed that classical computers using tensor network 
simulation could catch on current noisy quantum processors for random circuit sampling. While the linear cross 
entropy benchmark fidelity is used to certify all these claims, it may not capture in detail statistical 
properties of outputs. Here, we compare the bit strings sampled from classical computers using tensor 
network simulation by Pan {\it et al.} [Phys. Rev. Lett. 129, 090502 (2022)] and by Kalachev {\it et al.} 
[arXiv:2112.15083 (2021)] and from the Sycamore and Zuchongzhi quantum processors. It is shown that all
Kalachev {\it et al.}'s samples pass the NIST random number tests. The heat maps of bit strings show that 
Pan {\it et al.}'s and Kalachev {\it et al.}'s samples are quite different from the Sycamore or Zuzhongzhi 
samples. The analysis with the Marchenko-Pastur distribution and the Wasssertein distances demonstrates 
that Kalachev {\it et al.}'s samples are statistically close to the Sycamore samples than Pan {\it et al.}'s 
while the three datasets have similar values of the linear cross entropy fidelity. 
Our finding implies that further study is needed to certify or beat the claims of quantum advantage 
for random circuit sampling.
\end{abstract}
\maketitle

\section{Introduction}
The outperformance of quantum computers over classical computers for some computational tasks, called quantum 
advantage or quantum supremacy, is one of main subjects of quantum computing 
research~\cite{Feynman1982,Shor1997,HHL2009}. Random quantum circuit sampling~\cite{Boixo2018,Neill2018,Bouland2019} 
and Boson sampling~\cite{Scheel2004,Aaronson2011} are considered 
good computational tasks for demonstrating quantum advantage with current noisy quantum 
computers. In 2019, the Google team claimed the achievement of quantum advantage for 
random circuit sampling~\cite{Arute2019}. The Google Sycamore quantum processor with 53 qubits took 200 seconds 
to sample ten millions of bit strings from a random quantum circuit while a supercomputer at that time was 
estimated to take 10,000 years using the Schr\"odinger-Feynman algorithm to simulate the random quantum 
circuits~\cite{Arute2019}. Subsequently, the USTC team implemented random circuit sampling on the Zuchongzhi 
quantum processor with 56 and 61 qubits~\cite{Wu2021,Zhu2022_60qubit}. Also, quantum advantage for Boson 
sampling with optical qubits has been reported~\cite{Zhong2020,Zhong2021}. 

On the opposite side, to close the quantum advantage gap, the tensor network simulation of random 
quantum circuits has been performed on classical computers~\cite{Huang2020,Gray2021,Guo2021,Kalachev2021a,
Kalachev2021b,Pan2022a,Pan2022b}. For example, Kalachev~{\it et al.}~\cite{Kalachev2021a} reported 
that a tensor-contraction algorithm on a graphic processing unit took 14 days to sample millions of 
bit-string from the same random circuit that was implemented on the Sycamore processor with 53 qubits and 20 cycles. 
Similarly, Pan~{\it et al.}~\cite{Pan2022a,Pan2022b} showed that graphic processing units using a tensor 
network took 1 day to sample millions of bit-strings from the same random quantum circuit performed on 
the Sycamore processor with 53 qubits and 20 cycles. Thus, it is claimed that 
classical computers could catch up the current noisy quantum processors~\cite{Cho2022}.  

All of above arguments on quantum advantage for random circuit sampling on quantum or classical processors
are based on the two metrics: computational time and the linear cross-entropy benchmark fidelity 
(linear XEB). The first metric is just run time to accomplish a computational task and may be 
a primary measure to compare the performance of computers. The second metric is employed to ensure
that the output of noisy quantum processors is close to an ideal result.  
The linear XEB is zero if bit-strings are sampled from a classical uniform probability distribution. 
While a non-zero value of the linear XEB was used to support the claims of quantum advantage for
random circuit sampling, the liner XEB has some limitations in addition to the intrinsic scalability 
issue~\cite{Bouland2019}. The linear XEB could be spoofed without fully simulating a random quantum 
circuit~\cite{Gao2021,Barak2021}. Our previous works~\cite{Oh2022a,Oh2022b} demonstrated that
the linear XEB cannot capture the statistical properties of samples in the sense that
the Sycamore bit-strings are statistically different from the Zuchongzhi bit-strings while both processors 
show similar linear XEB values as a function of the number of qubits or the number of cycles. 

While the benchmark and certification of quantum processors are important to ensure the faithful 
calculation with a quantum processor, they are not fully explored~\cite{Eisert2020}. 
To validate the claim of quantum advantage and to verify quantum circuits, it would be better to 
compare directly the statistics of outcomes generated by quantum or classical processors. 
Quantum advantage regime means the verification with classical processors is very inefficient
In this paper, we compare bit 
strings sampled from the random circuits on the Sycamore processor~\cite{Martinis2022} and those 
calculated from tensor network simulation of the Sycamore random circuit on classical computers by 
Kalachev~{\it et al.}~\cite{Kalachev_data2022} and Pan~{\it et al.}'~\cite{Pan_data2022}. 
The paper is organized as follows. In Sec.~\ref{Sec:II}, we will perform the NIST random number 
tests on bit strings and plot the heat maps of bit strings. Also, we will calculate 
the Marchenko-Pastur distances and the Wasserstein distances between bit strings. These statistical 
analysis will show the samples from classical simulation are quite different from  the samples 
from the Sycamore quantum processor. In Sec.~\ref{Sec:III}, we will summarize our results 
and discuss the recent claim that quantum advantage for random circuit sampling is faded by 
classical computers. 

\section{Sampling from Random Circuits}
\label{Sec:II}

\subsection{Random Quantum Circuit Sampling}

Let us start with a brief summary of random quantum circuit sampling. The task of random quantum 
circuit sampling is to sample $n$-bit strings, $x \equiv a_1a_2\cdots a_n$, with $a_i = 0, 1$ 
and $i=1,\dots, n$ from a probability $p(x) = |\bra{x}U\ket{0^{\otimes n}}|^2$ defined by a random 
unitary operator $U$. Here, $\ket{x} = \ket{a_1a_2\cdots a_n}$ stands for a computational basis 
of $n$-qubits and $\ket{0^{\otimes n}} = \ket{00\cdots 0}$ for an initial state. In an ideal 
situation, a random unitary operator $U$ is sampled uniformly and randomly from the unitary group 
$\mathbb{U}(N)$ with $N=2^n$. A unique probability measure that is invariant under group multiplication 
is known the Haar measure. A set of random unitary matrices drawn from $\mathbb{U}(N)$ according to 
the Haar measure is called the circular unitary ensemble, which was introduced by Dyson~\cite{Dyson1962}. 

A random unitary matrix with respect to the Haar measure can be drawn
using the Euler angle method~\cite{Hurwitz1897,Zyczkowski1994,Diaconis2017} or the QR decomposition 
algorithm~\cite{Mezzadri2007}. Hurwitz~\cite{Hurwitz1897} showed that a unitary matrix 
can be constructed through the parameterization of $N^2$ elements with Euler angles. 
A random unitary matrix $U$ can be generated through the QR decomposition of
a matrix $Z$ with complex normal random entries, $Z=QR$ where $Q$ is a unitary matrix 
and $R$ is a upper-triangle matrix. Then, $U=Q\Lambda$ is a Haar-measure random unitary 
where $\Lambda$ is the diagonal matrix with diagonal 
entries $\Lambda_{ii} =R_{ii}/|R_{ii}|$ and $R_{ii}$ are the diagonal entries of the upper triangle matrix 
$R$~\cite{Mezzadri2007}. 

Emerson~{\it et al.}~\cite{Emerson2003} proposed the implementation of pseudo-random unitary operators 
on a quantum computer by applying single-qubit gates for each qubits and the simultaneous 
Ising interaction for all qubits. The Google and USTC experiments implemented a pseudo-random unitary 
operator that is composed of $m$ cycles of the sequence of $n$ single-qubit gates selected randomly from 
$\{\sqrt{X}, \sqrt{Y},\sqrt{W}\}$ and fixed two-qubit gates for according to tiling patterns.

A random unitary operator $U$ transforms the initial state $\ket{0^{\otimes n}}$ into a random 
quantum state 
\begin{align}
\ket{\psi} = U\ket{0^{\otimes n}} = \sum_{x=0}^{2^n-1} \sqrt{p_x}\,e^{i\theta_x} \ket{x} \,,
\end{align}
where $p_x = p(x)$. It is known that the probability distribution $Pr$ of $p_x$ with 
$x =0,\dots,2^n-1$ is given by $Pr(p) = (N-1) (1-p)^{N-2}$ with $N=2^n$,
the eigenvector distribution of the circular unitary ensemble~\cite{Dyson1962}.
For $N\gg 1$, it becomes $Pr(p)\approx N \exp(-Np)$ and is known as the $\chi^2$ distribution 
with 2 degrees of freedom, or the exponential distribution~\cite{Broady1981,Kus1988,Haake2010}. 
The output of random circuit sampling is a set of 
$M$ bit-strings, ${\cal D} = \{x_1,x_2,\dots,x_M\}$, generated by $M$ measurements 
of $\ket{\psi}$ in computational basis.
Note that a current quantum processor could generate only a bit-strings $x$, but not an 
amplitude $\sqrt{p_x}\,e^{i\theta_x}$, while a classical computer could calculate both quantities.
The phases $\theta_x$ are distributed uniformly in the range $[-\pi, \pi]$.

Today quantum processors are noisy and imperfect so the output of quantum computation could be
deviated from an ideal solution. It is important to verify whether quantum computation is performed
faithfully~\cite{Eisert2020}. Gilchrist~{\it et al.}~\cite{Gilchrist2005} suggested a golden standard for measures 
of the distance between ideal and noisy quantum processors. For sampling computation where the task 
is to sample outcomes $x$ from an ideal probability distribution $p(x)$, Gilchrist~{\it et al.} proposed
the the Kolmogorov distance $D(p,q)\equiv \sum_x |p(x) - q(x)|/2$ and the Bhattacharya overlap
$F(p,q) \equiv \sum_x \sqrt{p(x)q(x)}$ where $p(x)$ is an ideal distribution and $q(x)$ is
a real probability distribution generated by a quantum processor. On the other hand, Arute~{\it et al.} 
introduced the linear XEB 
\begin{align}
F_{\rm XEB} = \frac{2^n}{M}\sum_{i=1}^M p_{\rm ideal}(x_i) -1\,,
\label{Eq:XEB}
\end{align}
where the set of bit strings, ${\cal D} =\{x_1,x_2,\dots, x_M\}$, is the measurement outcomes 
of a noisy quantum processor and $p_{\rm ideal}(x) = |\bra{x}U\ket{0}^{\otimes n}|^2$ is an ideal 
probability of a quantum circuit $U$. For a classical uniform random distribution 
$p_{\rm ideal} = 1/2^n$, the linear XEB becomes zero. Basically, the linear XEB is derived from the cross 
entropy between the ideal distribution $p_{\rm ideal}(x)$ and the real distribution 
$p_{\rm real}(x)$ or the Kullback-Leibler divergence. However, the Kullback-Leibler 
divergence is not a true metric because it is not symmetric and does not satisfy the triangular 
inequality.

One of the main challenges in calculating the linear XEB or the Kolmogorov distance is how to obtain 
the ideal probability distribution $p_{\rm ideal}(x)$ or the real probability 
distribution $p_{\rm real}(x)$. For sampling calculation, a quantum computer produces bit-strings 
rather than $p(x)$. In quantum advantage regime, a classical computer is hard or inefficient 
to calculate $p(x)$ from quantum simulation. Also, it is impossible to estimate the empirical 
probability distribution from the measurement data of a quantum processor 
since the number of measurements $M$ is much less than $2^n$.
In the Google experiment with the Sycamore quantum processor, the number of measurements,
$M$, are order of $10^7$ but the dimension of the Hilbert space of 53 qubits is 
$2^{53}\sim 9\times 10^{15}$. 
Thus, to verify the performance of quantum computers in quantum advantage regime,
it would be better to compare directly two data sets, ${\cal D}_1 =\{x_1,x_2,\dots, x_M\}$ 
and ${\cal D}_2 =\{y_1,y_2,\dots, y_M\}$ without calculating the probability distributions 
$p(x)$ or $q(x)$. For example, since classical uniform random bit strings 
give rise to zero of the linear XEB, they could be one reference point. 

\subsection{Statistics of Samples of Random Circuits}
\subsubsection{Datasets of random circuit sampling}

We compare the 4 datasets of bit strings sampled from the Sycamore random circuits. The first is the 
Google dataset used in quantum supremacy experiments with the Sycamore quantum processor~\cite{Martinis2022}. 
In order to compare with other dataset, only sub-dataset for 53 qubits and cycle $m=12,14,16,18,20$
are considered. Each data file is identified using five parameters: $n$ is the number of qubits, $m$ the 
number of cycles $s$ seed for the pseudo-random number generator, $e$number of elided gates, and $p$ 
sequence of coupler activation patterns. For example, 
{\tt measurement}-{\tt n53}-{\tt m20}-{\tt s0}-{\tt e0}-{\tt pABCDCDAB.txt} stand 
for the measurement data file for 53 qubits, 20 cycles and {\tt ABCDCDAB} activation pattern.

The second dataset is Kalachev~{\it et al.}'s~\cite{Kalachev2021a}. They produced one million bit-strings 
for the Sycamore random circuit up to 20 cycles using the multi-tensor contraction algorithms and a modified 
frugal rejection sampling method.  To produce $k$ bit-strings, they first calculated $2k$ random batches and 
then selected $k$ bit-strings out of $2k$ batches using the frugal rejection sampling. We obtained 
Kalachev~{\it et al.}'s data from~\cite{Kalachev_data2022}. The data set is composed of 5 text files of 
one million bit 
strings for the Sycamore random circuit with $m= 12, 14, 16, 18$, and $20$ cycles (for example, 
{\tt samples}-{\tt m20}-{\tt f0}-{\tt 002.txt}). For cycle $m = 12, 14, 16, 18$, the target fidelity is 0.02 and 
for $m=20$, 
the target fidelity is 0.02. Also, one text file for spoofing the linear XEB ({\tt spoofing}-{\tt m20.txt}) 
is provided. 

The third dataset is generated by Pan {\it et al}~\cite{Pan_data2022} using the tensor network simulation. 
However, only the spoofing dataset with about one million bit strings, {\tt samples-metropolis.txt}, 
is available. Finally, we prepare the file of bit strings sampled from a classical uniform random 
distribution ({\tt n50}-{\tt classical.txt}). Note that the Zuchongzhi data for $n=53$ qubit are not available
~\cite{Wu2021} so we compare only the Sycamore dataset with Kalachev {\it et al.}'s data and 
Pan {\it et al.} data.

\subsubsection{NIST random number tests.}

If one take a close look at bit strings sampled from a random quantum circuit, bit 0 and 1 seem 
to appear randomly. Note that we are focusing on randomness of bits of a sample, 
not bit-strings. As discussed, the probability of finding a bit string $x$ is given by 
$p(x)= |\bra{x}U\ket{0^{\otimes n}}|^2$ and is not uniformly random. So some bit-strings are 
sampled more likely.  While it is not clear whether a random circuit produces bit 0 or 1 uniformly 
at random, our previous studies~\cite{Oh2022a,Oh2022b} showed that some of the Zuchongzhi samples 
pass the NIST random number tests~\cite{NIST2010}. Recently, Shi {\it et al.}~\cite{Shi2022} 
demonstrated a possibility of using the Boson sampling method to generate true random bits. 
They showed that bit-strings generated from the Boson sampling on optical qubits pass the NIST 
random number tests while more simple ways of generating true quantum random numbers 
exist~\cite{Collantes2017}. Here, we perform the NIST random number tests~\cite{NIST2010,Ang2019} 
on bit strings sampled form the Sycamore random circuit using tensor network simulation: Kalachev
{\it et al.}'s five samples and Pan{\it et al.}'s spoofing sample. As expected, Pan {\it et al}'s 
spoofing sample does not pass the NIST random number tests because first several bits of 56-bit 
strings are fixed to zeros or ones. We find that all five samples of Kalachev {\it et al.} 
pass the NIST random number tests. The details of the NIST random number tests for the six 
samples are shown in Supplemental Material~\cite{Suppl2022}. However, we leave an open question 
whether bits of bit-strings sampled from a random circuit should be uniformly at random.

\begin{figure*}[t]
\includegraphics[width=0.245\textwidth]{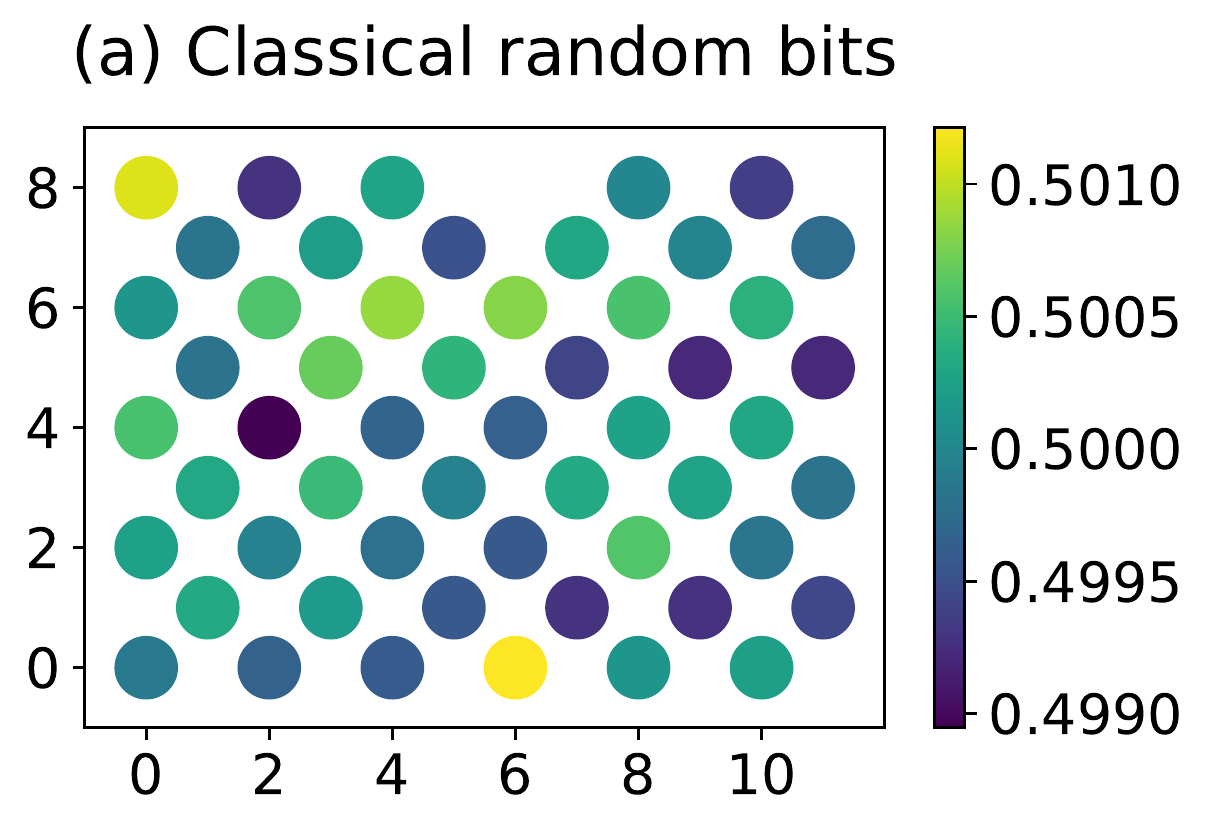}
\includegraphics[width=0.245\textwidth]{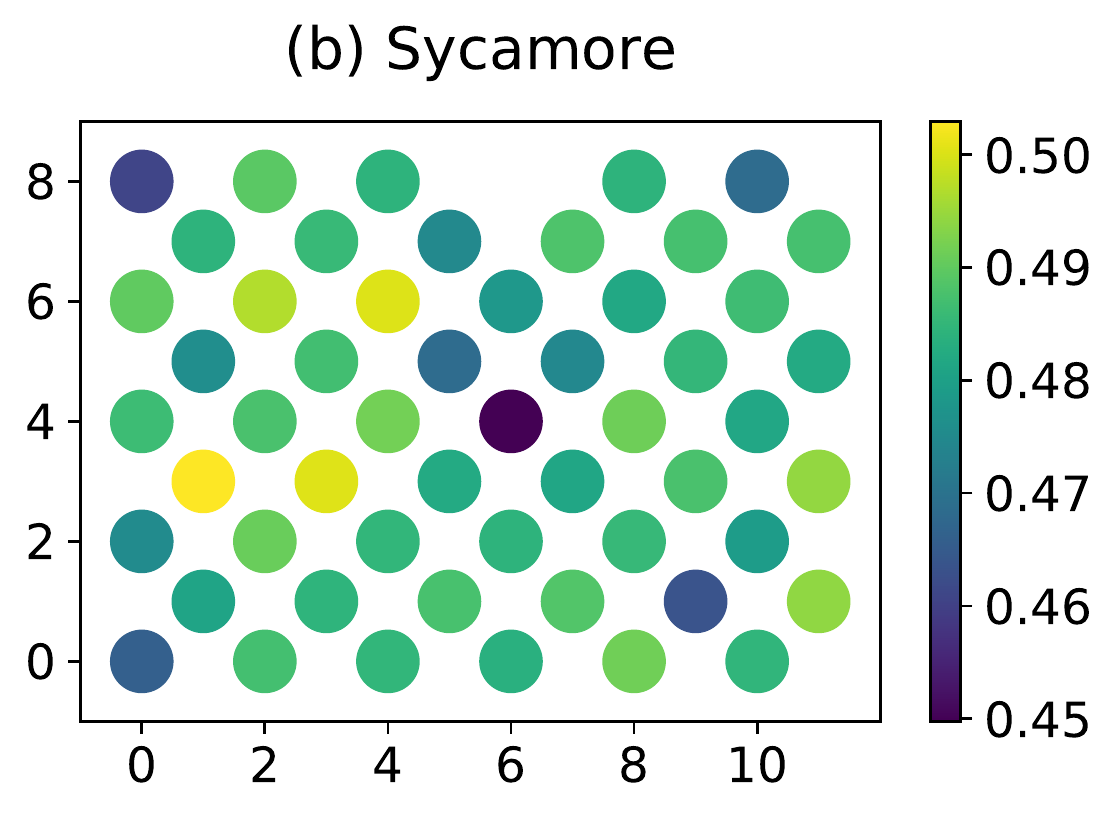}
\includegraphics[width=0.245\textwidth]{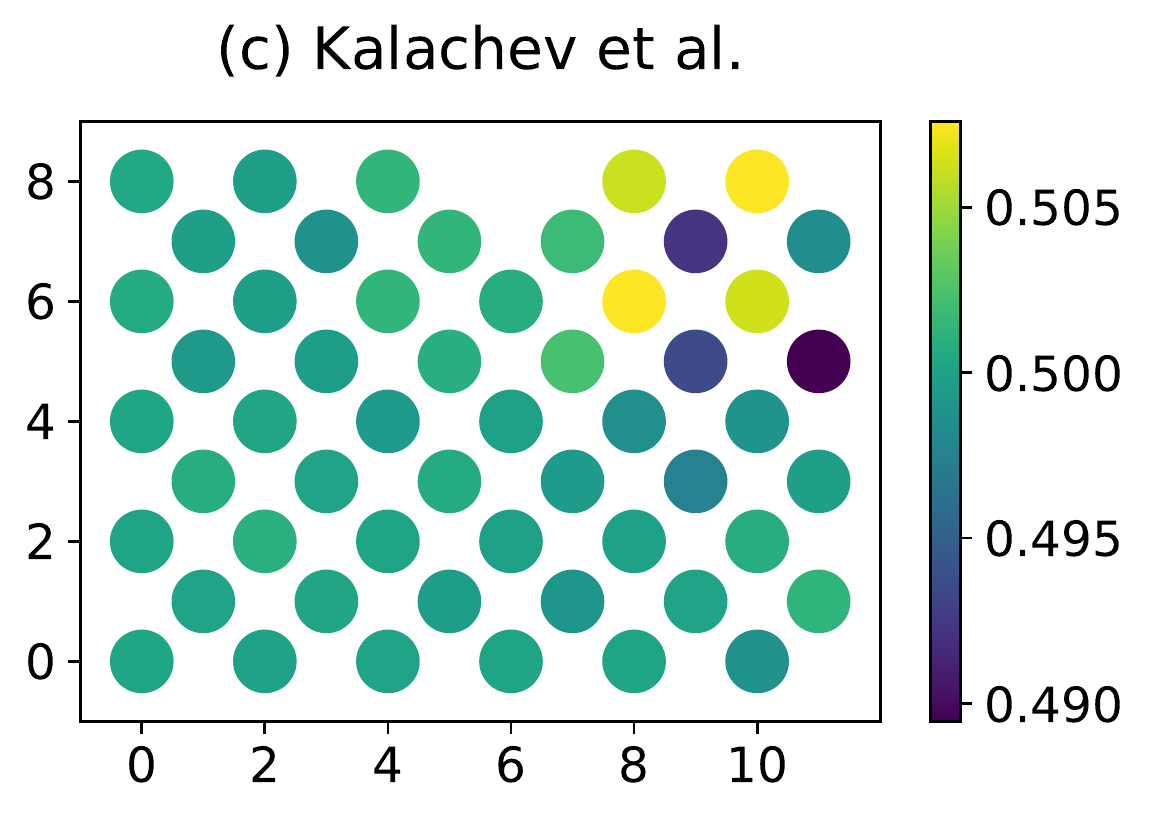}
\includegraphics[width=0.245\textwidth]{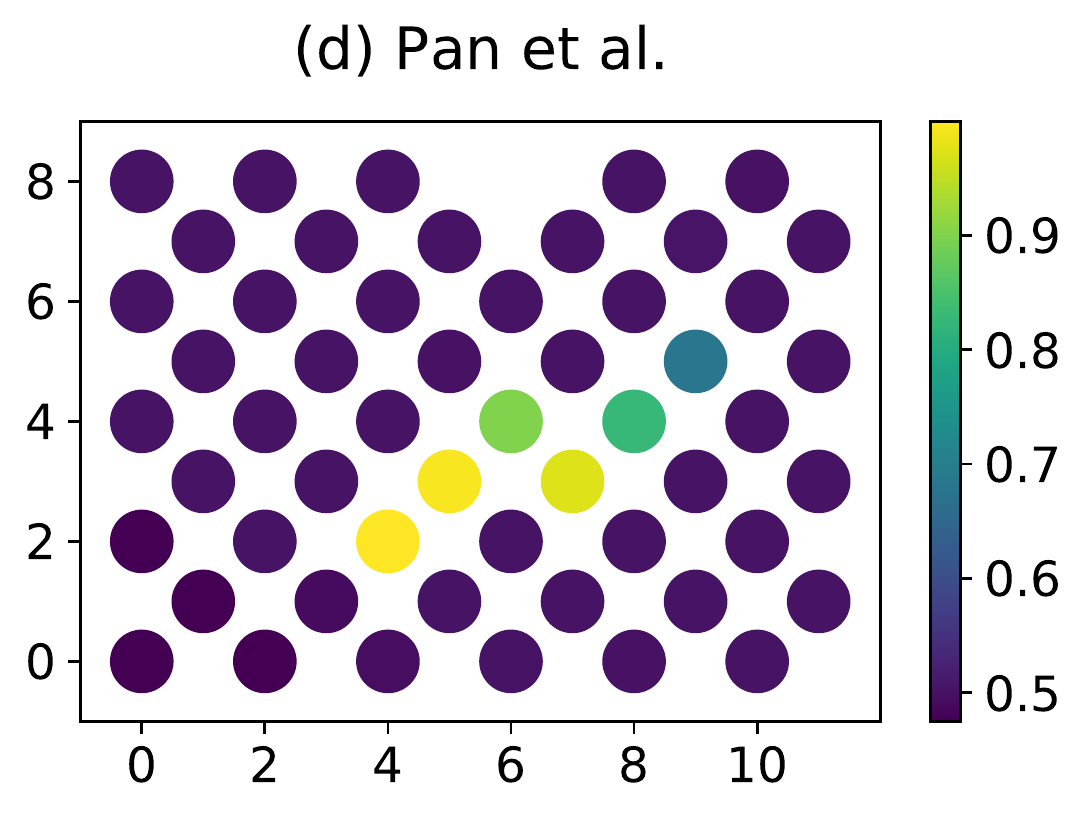}\\[11pt]
\includegraphics[width=0.245\textwidth]{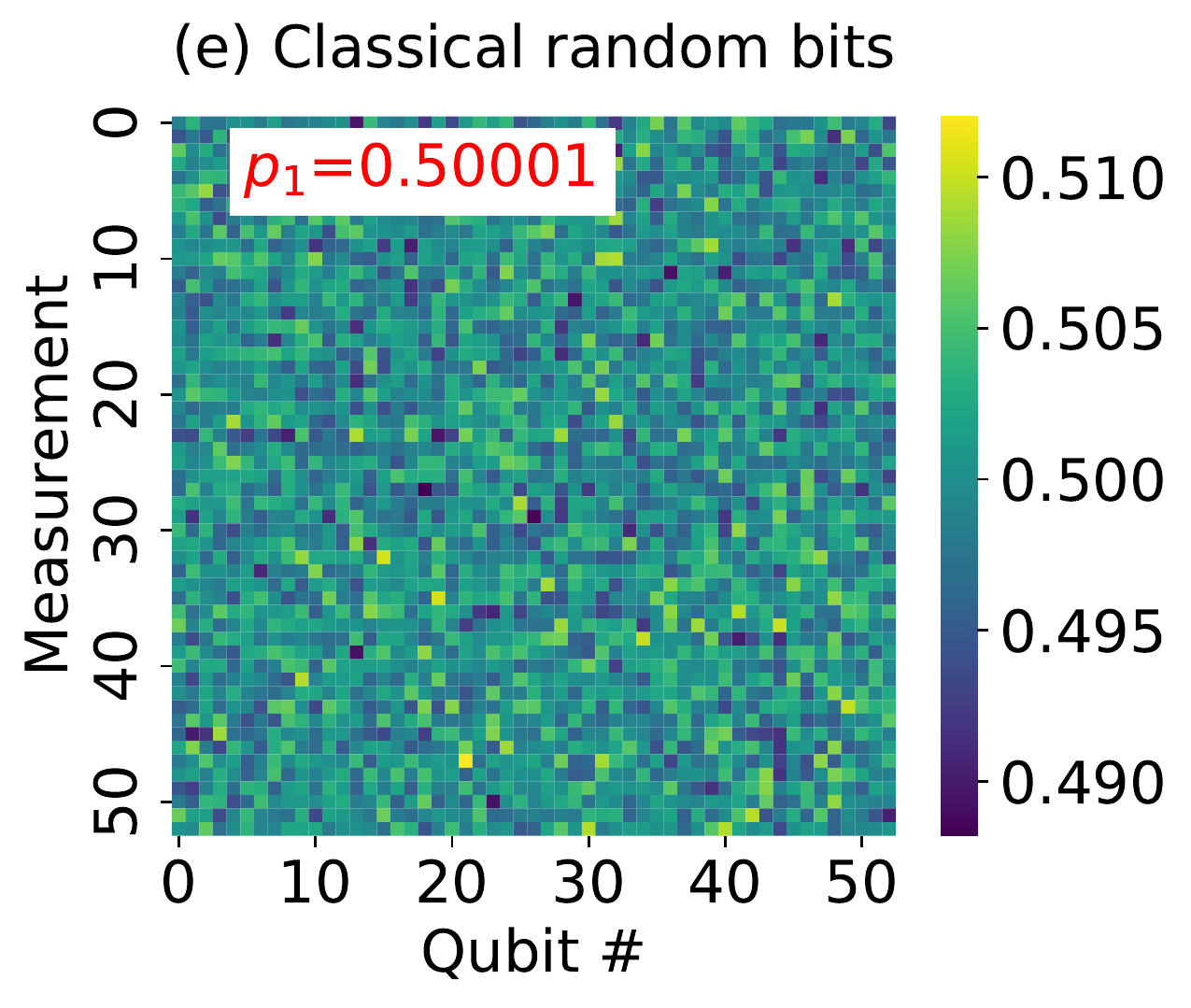}
\includegraphics[width=0.245\textwidth]{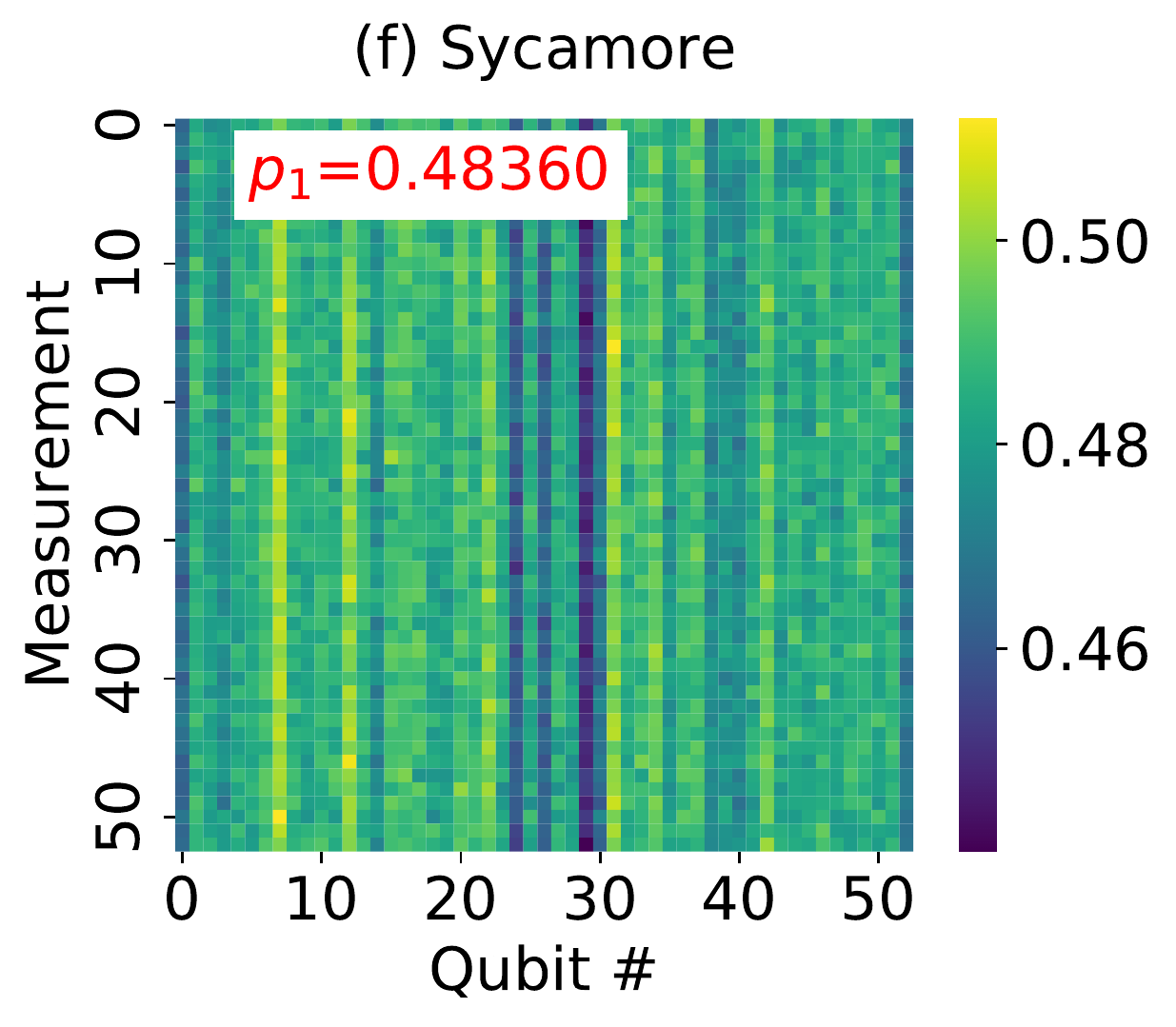}
\includegraphics[width=0.245\textwidth]{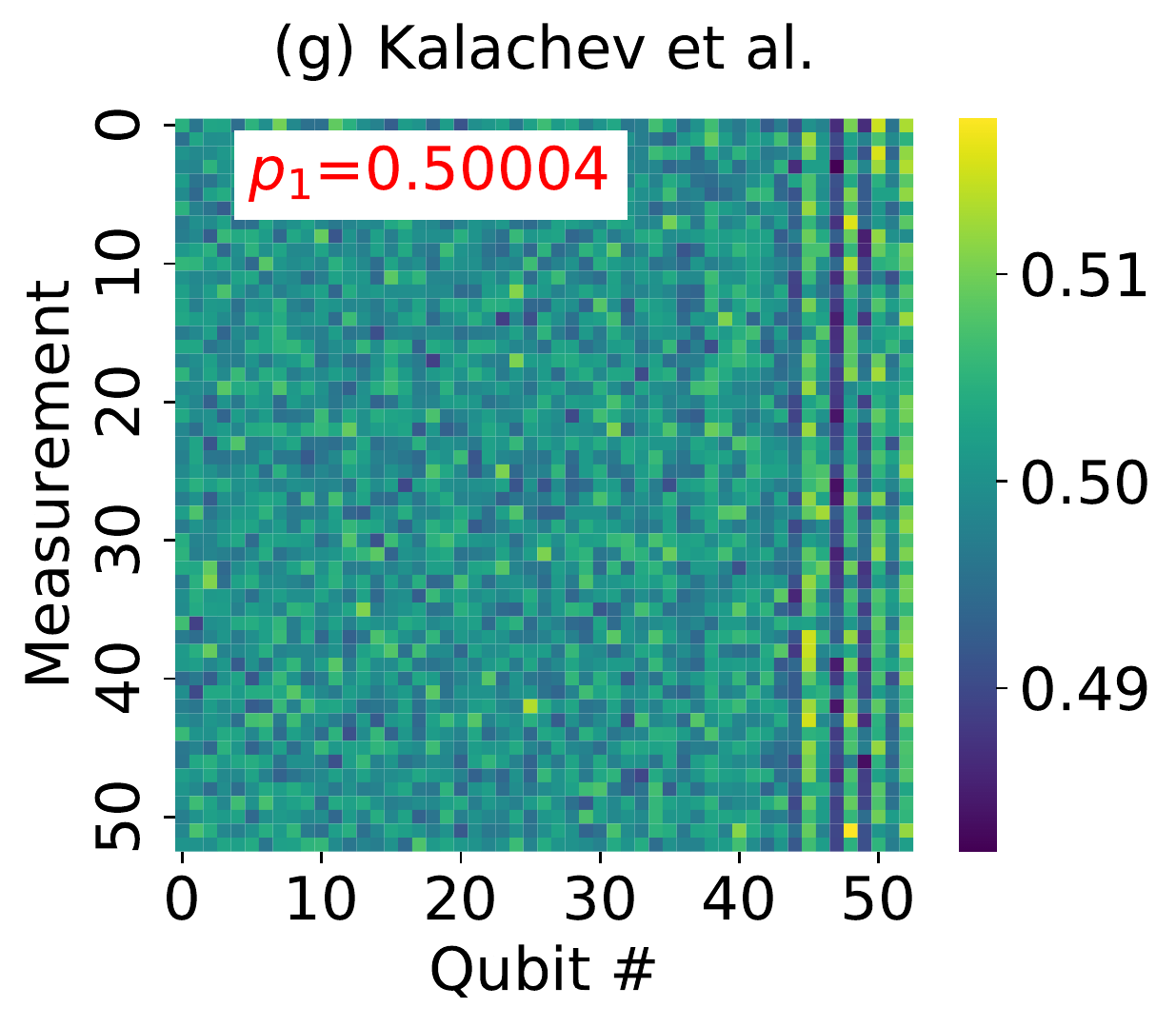}
\includegraphics[width=0.245\textwidth]{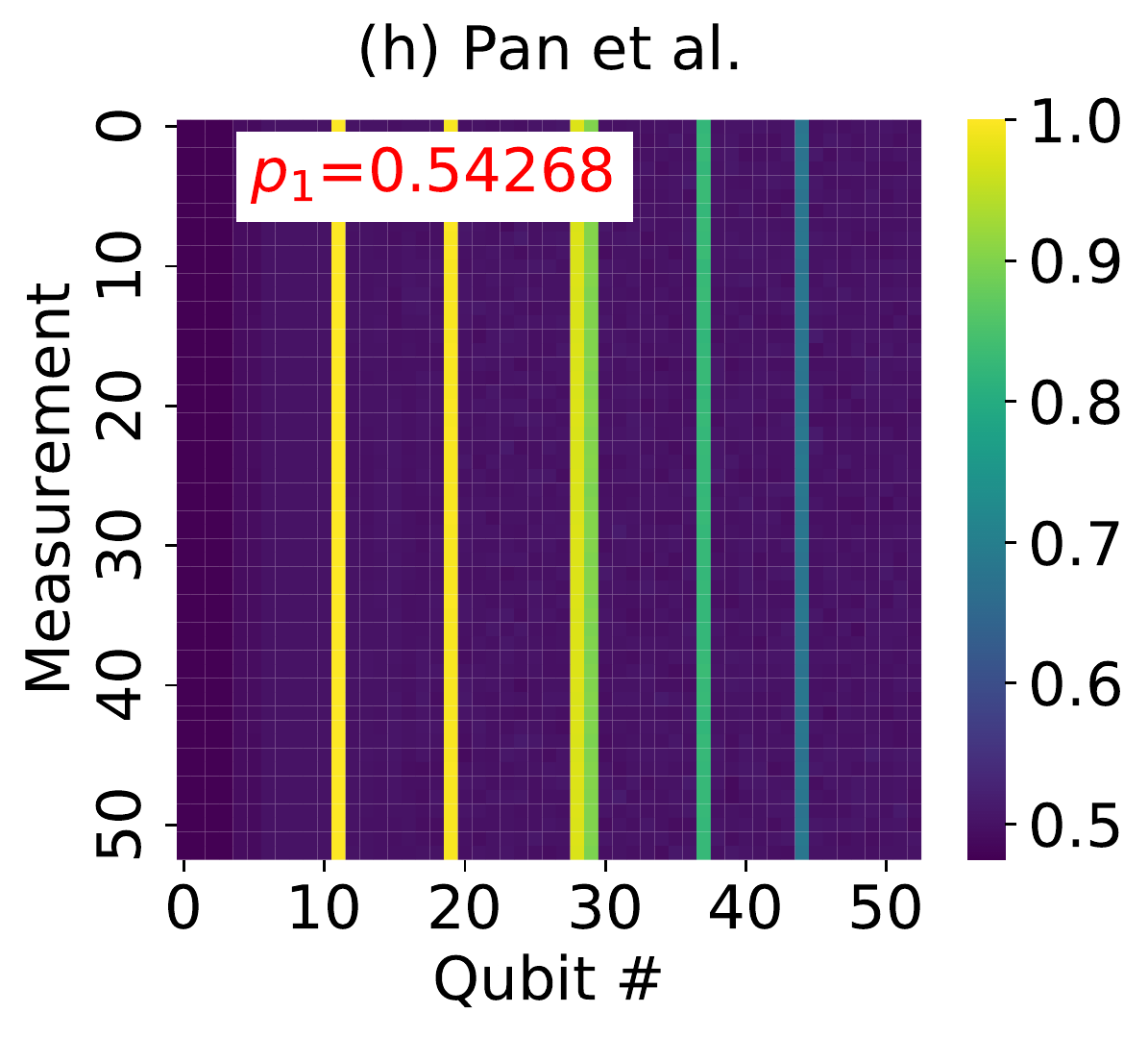}
\caption{\label{Fig1} (First row) For 2-dimensional arrays of the Sycamore quantum processor with 53 qubits,
the averages of finding outcome 1 of each qubit over one million measurements are plotted for 
(a) the classical random bit strings, (b) the Sycamore sample 
({\tt measurement}-{\tt n53}-{\tt m20}-{\tt s0}-{\tt e0}-{\tt pABCDCDAB.txt}), 
(c) Kalachev {\it et al.} sample ({\tt samples}-{\tt m20}-{\tt f0}-{\tt 002.txt}), and (d) Pan sample 
({\tt sample}-{\tt metropolis.txt}). (Second row) The heat maps are plotted for (e) classical random bits
(f) the Sycamore sample, (f) Kalachev {\it et al.}'s sample, (g) Pan {\it et al.}'s sample.
The average of finding bit 1 over the sample is denoted by $p_1$.
}
\end{figure*}

\subsubsection{Heat maps of bit strings.}

A heat map is a good way of visualizing the statistical properties of outcomes of each qubits. 
The averages of outcomes at each qubit arranged in the 2-dimensional array of the Sycamore processor 
are plotted for the classical random bits in Fig.~\ref{Fig1}~{(a)}, the Sycamore sample in 
Fig.~\ref{Fig1} (b), the Kalachev sample in Fig.~\ref{Fig1} (c), and the Pan {\it et al.}'s 
sample in Fig.~\ref{Fig1} (d), Kalachev {\it et al.}'s sample and Pan {\it et al.}'s sample, both of which were 
obtained using the tensor network simulation, look like only that a random quantum circuit acts on only some 
parts of 53 qubits. Especially, Pan {\it et al}'s spoofing sample is quite different from the other 
three samples. As the animations in Supplemental Material show~\cite{Suppl2022}, for Pan {\it et al'}
spoofing sample, some qubits are turned always off (and turned on later). 

To see the statistical difference among samples in detail, we slice a $10^6\times 53$ bit array, 
${\cal D} = (x_1,x_2,\dots,x_M)^t$, into $53\times 53$ square arrays $\{A^{(i)}\}$ with 
$i=1,\dots,L$ and $L\equiv{\rm int}(M/53)$. The heat maps of the averages of $A^{(i)}$, 
$(1/L)\sum_i A^{(i)}$, for the classical random bit sample, the Sycamore sample, the Kalachev {\it et al.}'
sample, and the Pan {\it et al.}'s sample are plotted in Figs.~\ref{Fig1}~{(e)}, (f), (g), and (h), 
respectively. The heat map of the classical random bits has no stripe patterns as expected. 
The heat map of Pan {\it et al.} 
sample exhibits also the stripe patterns, but is quite different from those of the Sycamore 
sample of Kalachev {\it et al.}'s. It is interesting to see the heat map 
of Kalachev {\it et al}'s samples, Fig.~\ref{Fig1}~{(f)} shows stripe patterns similar to 
that of the Sycamore sample, Fig.~\ref{Fig1}~{(e)} while the former passes the NIST random number tests. 
One may expect that bit strings which pass the NIST random number tests would do not show any patterns. 
However, Kalachev {\it et al}'s samples here and Zuchongzhi samples in our previous work~\cite{Oh2022b} 
may break a common belief on random bit or random numbers.

We count the number of outcome $1$ of a sample composed of one million bit-strings, denoted by $p_1$.
As shown in Figs.~\ref{Fig1} (e), (f), (g) and (h), the averages of finding outcome 1 are estimated 
as $p_1=0.50001$ for the classical random bit sample, $p_1=0.48360$ for the Sycamore sample,
$p_1=0.50004$ for Kalachev {\it et al}'s sample, and $p_1=0.545000$ for Pan {\it et al}'s sample.
Note that the averages of finding outcome 1 for Zuchongzhi samples are almost 0.5.
the Sycamore samples have $p_1$ less than 0.5 because of the readout errors~\cite{Rinott2020}.

%
\subsubsection{Marchenko-Pastur distances and Wasserstein distances}

One expects that two classical processors will produce the same output or statistically similar outputs. 
Like that, two quantum processors implement similar quantum gates will produce similar 
outcomes~\cite{Nielsen_Chuang}. Suppose a unitary operator $V$ of an ideal processor is approximate to 
a unitary operators $U$ of a real processor. The error between $U$ and $V$ is defined 
by $E(U,V) \equiv \max_{\ket{\psi}} ||(U-V)\ket{\psi}||$. If the difference in two gates is small,
then measurement statistics of $U\ket{\psi}$ is approximated by $V\ket{\psi}$. This is written
as the inequality~\cite{Nielsen_Chuang} 
\begin{align}
|p_U -p_V| \le 2 E(U,V)\,.
\end{align}
If the error $E(U,V)$ is small, the difference in probabilities between measurement outcomes, $p_U$
and $p_V$, is also small.

\begin{figure*}[t]
\includegraphics[width=0.33\textwidth]{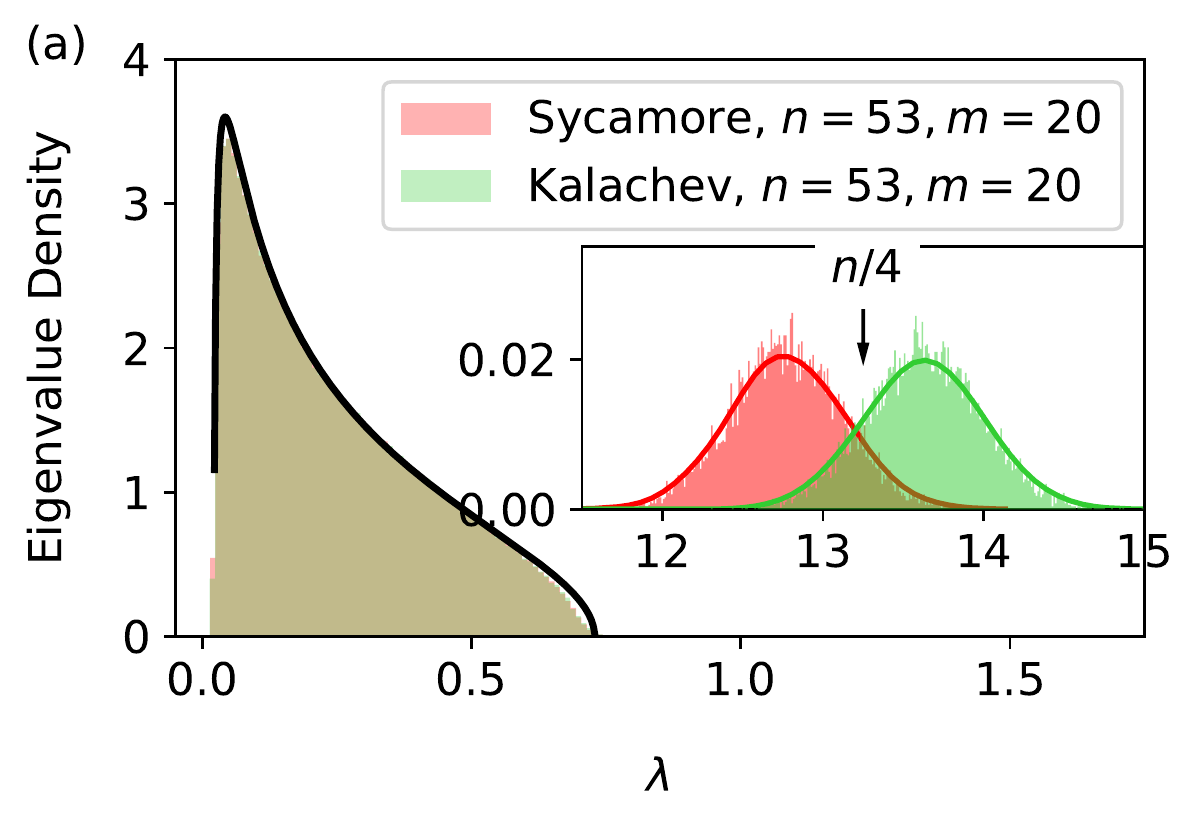}
\includegraphics[width=0.33\textwidth]{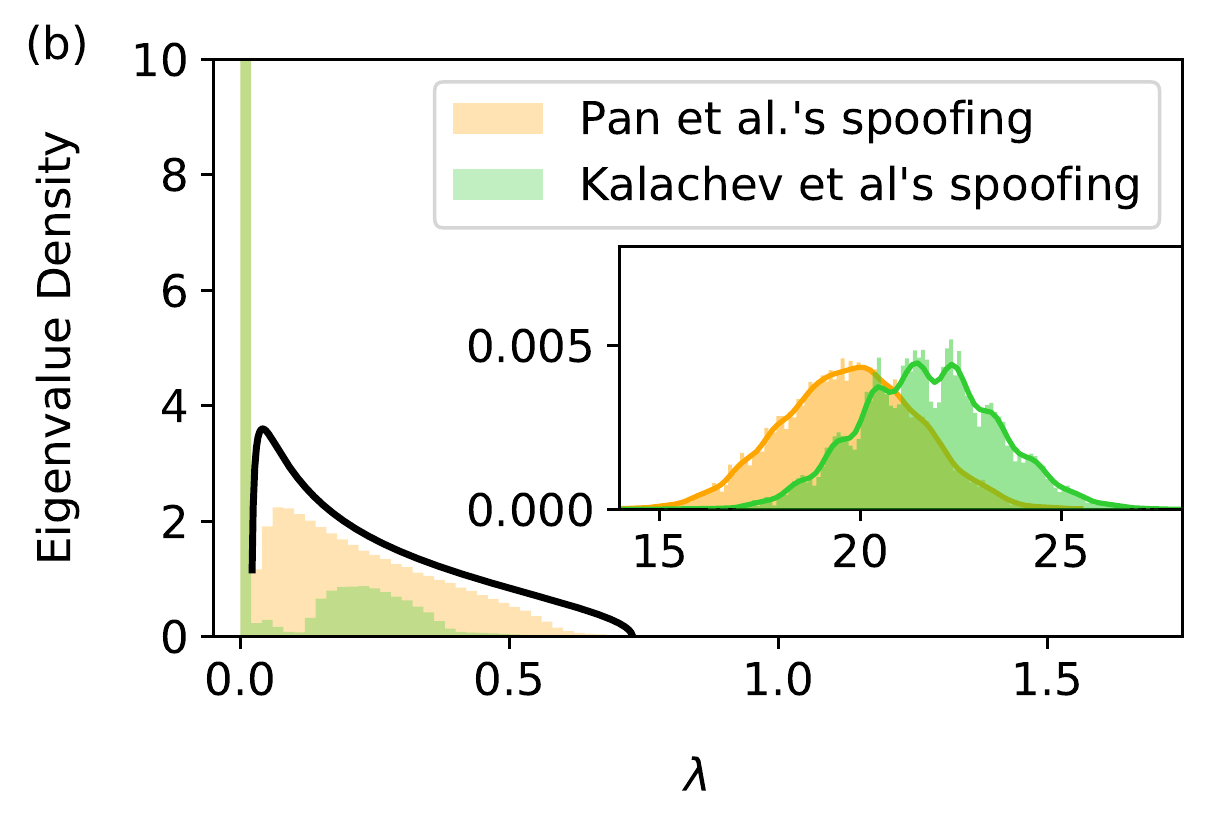}
\includegraphics[width=0.31\textwidth]{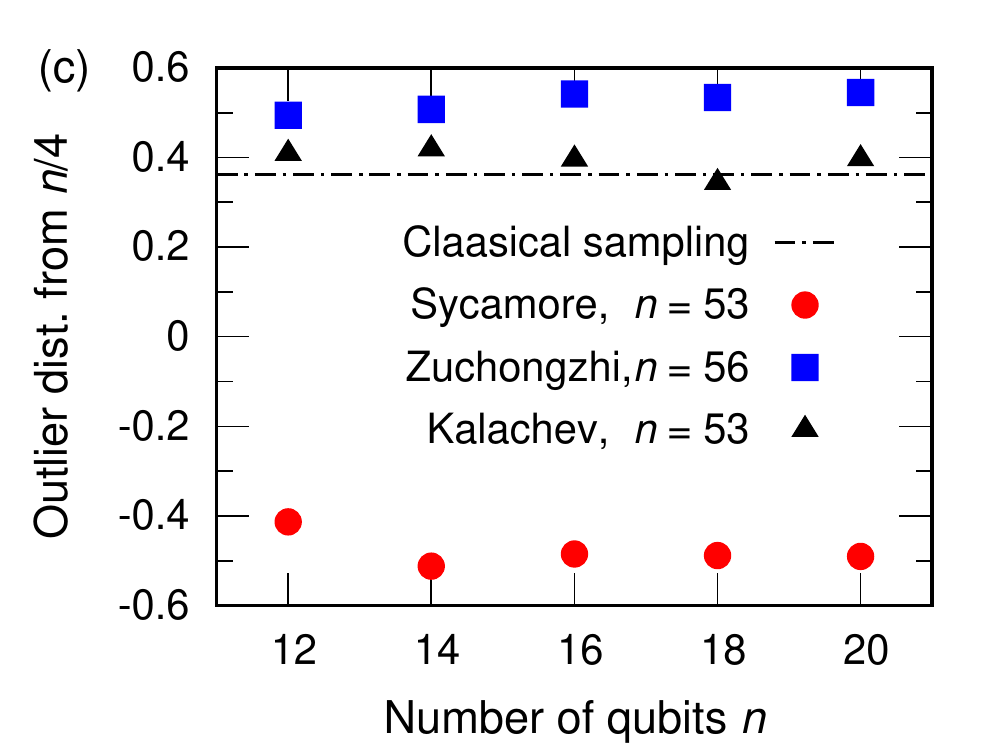}
\caption{\label{Fig2} 
(a) Marchenko-Pastur distribution of eigenvalues of $(1/k)X^t\cdot X$, Eq.~(\ref{Eq:MP}), 
for the Sycamore sample
with $n=53$ and $m=20$ (red color) and for the Kalachev {\it et al.}'s sample with $n=53$ and $m=20$ 
(orange color). The inset shows the outlier peaks around $n/4$.
(b) Marchenko-Pastur distribution of eigenvalues of $(1/k)X^t\cdot X$ for Pan {\it et al.}'s
spoofing sample ({\tt samples}-{\tt metropolis.txt}) and Kalachev {\it et al.}'s spoofing sample 
({\tt spoofing}-{\tt m20.txt}). The inset shows the outliers.
(c) Marchenko-Pastur distances (outlier distances from $n/4$) as a function of the number of cycles 
$m$ for the Sycamore samples and Kalachev {\it et al.}'s samples with $n=53$ qubits, and 
for the Zuchongzhi samples with $n=56$.
}
\end{figure*}

If two processors, no matter processor are quantum or classical, implement the similar unitary operators 
for random circuit sampling, statistical distances between probabilities should be close to each other. 
In practice, the two probability distributions, $p_U(x)$ and $p_V(y)$, may be unknown or unavailable,
but only outcomes ${\cal D}_U = \{x_1,x_2,\dots,x_M\}$ and ${\cal D}_V = \{y_1,y_2,\dots,y_M\}$
sampled from $p_U(x)$ and $p_V(y)$ are available. Thus one needs to calculate the statistical distances 
between two samples, ${\cal D}_U$ and ${\cal D}_V$. To this end, we employ the Marchenko-Pastur distances 
and the Wasserstein distances between samples.

The linear XEB, Eq.~(\ref{Eq:XEB}) is zero if bit strings are sampled from classical uniform random bits and 
a sample generated from a random circuit would be different. Random matrix theory, here the 
Marchenko-Pastur distribution of random matrices~\cite{MarPas67}, could be helpful in capturing this difference.
An $M\times n$ binary array ${\cal D} = (x_1, x_2,\dots, ,x_M )^t$ is sliced into
into the collection of $k \times n$ rectangular random matrices $X$. Here we set $k=2n$.
If all entries of $X$ are independent and identically binary random variables 
$\{0, 1\}$, then one has the mean $\mu_X = 1/2$ and the variance $\sigma_X^2 = 1/4$. 
We calculate the distribution of eigenvalues of square matrices $(1/k)X^t\cdot X$ as shown in 
Fig.~\ref{Fig2}. The eigenvalue distribution of $X$ is composed of the two parts:
the bulk distribution is the Marchenko-Pastur distribution corresponding to random noise
and the outlier represents the signal~\cite{Oh2022a,Oh2022b}. 
The random matrices $X$ can be written as $Y = 2X-J$ where $J$ of all entries 1, and $Y$ has the zero 
mean $\mu_Y =0$ and the variance $\sigma^2_Y =1$. The square random matrix $X$ can be expressed as
\begin{align}
\frac{1}{k}X^t\cdot X = \frac{1}{4k}\left(Y^t\cdot Y + X^t\cdot J + X^t\cdot Z + J^t\cdot J \right)\,,
\label{Eq:XtX}
\end{align}
Here, the eigenvalue distribution of the first term in Eq.~(\ref{Eq:XtX}) is given by the Marchenko-Pastur 
distribution~\cite{MarPas67} 
\begin{align}
\rho(\lambda) = \frac{1}{2\pi\sigma^2 \gamma} \frac{\sqrt{(\lambda_+ -\lambda)(\lambda -\lambda_{-})}}{\lambda}\,,
\label{Eq:MP}
\end{align}
where $\gamma = n/k$ is the rectangular ratio and $\lambda_{\pm} = \sigma^2(1 \pm\sqrt{\gamma})^2$ are the upper
and lower bounds. For $k=2n$, one has $\gamma = 1/2$ and $\lambda_+ = 1 +\sqrt{1/2}$. 
The eigenvalues of the last term $(1/4k) J^t\cdot J$ are 0 and $n/4$. This is the location of 
the outlier.  We call {\it the Marchenko-Pastur distance} the distance of the peak position of the outlier 
of a sample from $n/4$.

Fig.~\ref{Fig2} (a) plots the Marchenko-Pastur distribution of eigenvalues of the ensemble of $(1/2n){X^t\cdot X}$
for the Sycamore sample ({\tt measurement}-{\tt 53}-{\tt m20}-{\tt s0}-{\tt e0}-{\tt pABCDCDAB.txt}) 
and the Kalachev {\it et al.}'s sample ({\tt samples}-{\tt m20}-{\tt f0}-{\tt 002.txt}). The majority of 
the eigenvalues follow the Marchenko-Pastur distribution given by Eq.~(\ref{Eq:MP}). The inset of
Fig.~\ref{Fig2} (a) shows the outliers of the distribution of eigenvalues around $n/4$, 
here $53/4 =13.25$ for $n=53$ qubits. The distance of the peak of the outliers from $n/4$ is called
the Marchenko-Pastur distance which is used to measure how a sample is deviated from classical random bit 
strings. Fig~\ref{Fig2} (b) plots the Marchenko-Pastur distribution for Pan {\it et al.}'s spoofing sample
({\tt samples}-{\tt metropolis.txt}) and Kalachev {\it et al.} spoofing file ({\tt spoofing}-{\tt m20.txt}).
As expected, the eigenvalue distributions of both spoofing samples are quite different from those of the Sycamore
samples or those of Kalachev {\it et al.}'s samples. There is the big peak at eigenvalue zero, $\lambda =0$.
Also, the outliers are far away from $n/4$.

Fig.~\ref{Fig2}~(c) plot the Marchenko-Pastur distance of samples from $n/4$ as a function of the number of 
cycles $m=12,14,16,18,$ and $20$. As shown in Fig.~\ref{Fig2}~(b), the Marchenko-Pastur distances of 
Kalachev {\it et al.}'s samples are very close to that of the classical random bit strings, approximately 0.36.
The closeness of Kalachev {\it et al.}'s samples to the classical random bit strings are consistent with
the fact that Kalachev {\it et al.}'s samples pass the NIST random number tests discussed above.
We could not calculate the Marchenko-Pastur distribution for Pan {\it et al.}'s sample because

\begin{figure}[t]
\includegraphics[width=0.35\textwidth]{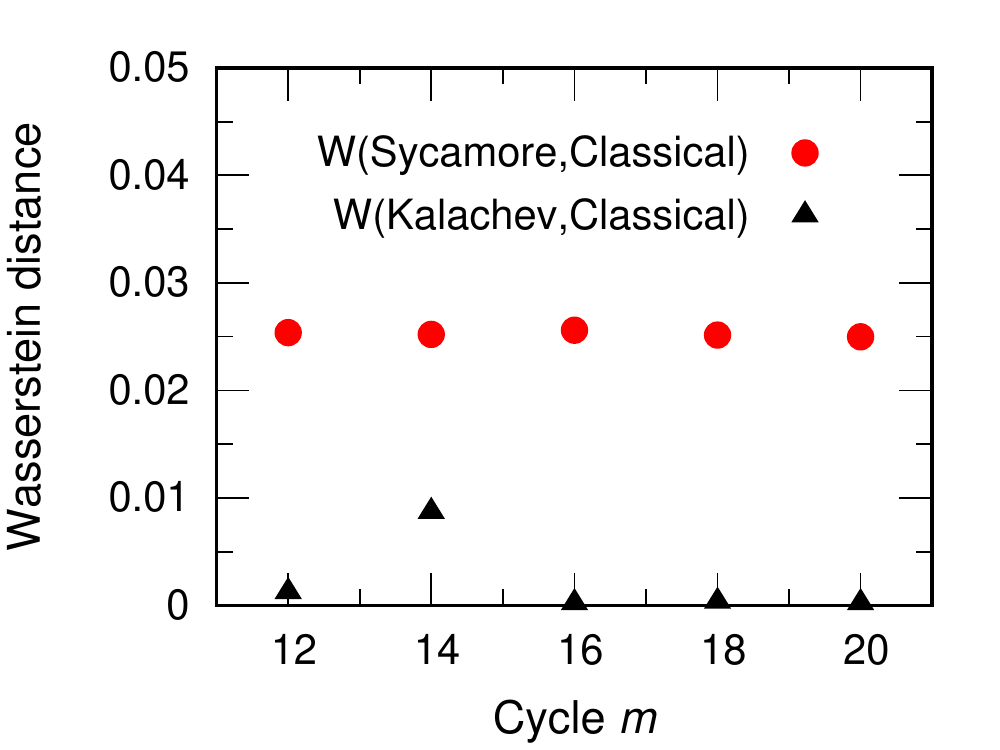}
\caption{\label{Fig3}
Wasserstein-distances of the Sycamore samples qubits with $n=53$ and 
the Kalachev {\it et al.}'s samples with $n=53$ from the classical random bit sample. 
}
\end{figure}

In order to support the analysis above using the Marchenko-Pastur distances between samples, we calculate
the Wasserstein distances for samples~\cite{Villani2008}. In contrast to the Kullback-Leibler divergence, 
the Wasserstein distance is a metric on probability distributions. It is symmetric and satisfies 
the triangular inequality. The $\alpha$-Wasserstein distance between two probability distribution $p(x)$ 
and $q(y)$ is given by~\cite{Villani2008}
\begin{align}
W(p,q) = \left( \inf_{\pi\in \Gamma(p,q)} 
\int_{\mathbb{R}\times\mathbb{R}}  |x - y|^\alpha\, d\pi(x,y) \right)^{1/\alpha}\,,
\end{align}
where $\Gamma(p,q)$ is the set of joint distributions $\pi(x,y)$ 
whose marginals are $p(x)$ and $q(y)$.
If $p(x)$ is the empirical distribution of a dataset $D_1 = \{x_1,x_2,\dots,x_M\}$ and 
$q(y)$ is the empirical distribution of a dataset $D_2 = \{y_1,y_2,\dots,y_M\}$, then
$\alpha$-Wasserstein distance is written as
\begin{align}
W_\alpha(p,q) =  \left(\sum_{i=1}^M |x_{(i)} -y_{(i)}|^\alpha \right)^{1/\alpha}\,,
\end{align}
where $x_{(k)}$ denotes the order statistic of rank $k$, i.e., $k$-th smallest value in the dataset ${\cal D}_1$.
Here, we calculate the first Wasserstein distance for the datasets of random circuit sampling using   
the python optimal transport library~\cite{Flamary2021pot}.
To this end, each bit string $x$ is converted to a value in range $[0,1]$ by dividing it by $1/2^n$.
Also, we consider only 1 million bit strings. Fig.~\ref{Fig3} plots the Wasserstein distances of samples
from the classical random bit strings as a function of the number of cycles $m$. It is shown that 
the Wasserstein distances between Kalachev {\it et al.}'s samples and the classical random bit strings
are very close. This result is consistent with the analysis with the Marchenko-Pastur distance.
The Wasserstein distances of Pan {\it et al}'s spoofing sample from the classical random sample,
the Sycamore sample, Kalachev {\it et al.}'s spoofing sample are 
about 0.0234, 0.01390, and  0.3275, respectively.

\section{Conclusion}
\label{Sec:III}

In summary, we have investigated the statistical properties of bit-strings sampled from random circuits
on the Sycamore quantum processor and obtained using the tensor network simulation on classical processors. 
We considered Kalachev {\it et al.}'s samples~\cite{Kalachev2021a,Kalachev2021b,Kalachev_data2022} 
and Pan {\it et al.}'s spoofing sample~\cite{Pan2022a,Pan2022b,Pan_data2022}, and compared them with
the Sycamore dataset~\cite{Martinis2022} and the classical random bit strings. We found that Kalachev
{\it et al.}'s samples pass the NIST random number tests while they have stripe patterns of 
the heat maps. The heat maps of the two spoofing samples were shown to be quite different from 
those of the Sycamore samples. Since some parts of bit-strings of the spoofing samples are fixed 
to zeros or ones, the outcomes of the spoofing samples look like that a random circuit applied only 
a few qubits. So one can easily distinguish them from the Sycamore samples while the linear XEB values of
the spoofing samples are greater than zeros. Using the Marchenko-Pastur distribution of eigenvalues
of the samples and the Wasserstein distances between the samples show that Kalachev {\it et al.}'s 
sample is close to classical random bit strings and the spoofing samples are far away from the classical
samples or from the Sycamore samples. 

Google's 2019 experiment on random circuit sampling~\cite{Arute2019} was a landmark in quantum computing and 
has been much attraction as well as intense debates~\cite{Kalai2022}. The tensor network simulation on 
quantum circuits has been proven to be effective compared the Schr\"odinger-Feynman simulation and could catch up
with current noisy quantum processors~\cite{Kalachev2021a,Kalachev2021b,Pan2022a,Pan2022b,Cho2022}.
All claims are based on two metrics: computational time and the linear XEB. While the linear XEB may serve 
as a good measure to verify quantum advantage, its limitation has been pointed 
out~\cite{Gao2021,Kalachev2021a,Pan2022b}. Other measures could be used to verify quantum computation 
for sampling~\cite{Gilchrist2005}. For example, the random matrix analysis or the Wasserstein distances
~\cite{Oh2022a,Oh2022b} would tell us other aspects of outcomes of random circuit sampling.
As more results from quantum processors and advanced simulation on classical processors become available,
a comparative study could deepen our understanding of quantum advantage.
Our results raise a question about the claim 
that quantum advantage for random circuit sampling is faded by classical 
computers~\cite{Pan2022b,Cho2022}. 

\begin{acknowledgments}
This material is based upon work supported by the U.S. Department of Energy, Office of Science, National 
Quantum Information Science Research Centers. We also acknowledge the National Science Foundation under 
Award number 1955907. 
\end{acknowledgments}

\bibliography{rqc}
\vfill
\end{document}